\documentclass[transmag]{IEEEtran}
\usepackage{latexsym}
\usepackage{graphicx}
\usepackage{amsfonts,amssymb,amsmath}
\usepackage{hyperref}
\def\BibTeX{{\rm B\kern-.05em{\sc i\kern-.025em b}\kern-.08em T\kern-.1667em\lower.7ex\hbox{E}\kern-.125emX}}
\begin{document}

\title{Unidirectional Curved Surface Plasmon
Polariton in a Radially Magnetized System}

\author{S. Pakniyat, \IEEEmembership{} A.H. Holmes, \IEEEmembership{Student, IEEE}, G.W. Hanson, \IEEEmembership{Fellow, IEEE}

\thanks{Funding for this research was provided by the National Science Foundation
under grant number EFMA-1741673.}
\thanks{Department of Electrical Engineering, University of Wisconsin-Milwaukee,
Milwaukee, Wisconsin 53211, USA (pakniyat@uwm.edu)}
\thanks{Department of Electrical Engineering, University of Wisconsin-Milwaukee,
Milwaukee, Wisconsin 53211, USA (holmesam@uwm.edu).}
\thanks{Department of Electrical Engineering, University of Wisconsin-Milwaukee,
Milwaukee, Wisconsin 53211, USA (george@uwm.edu).}}

\IEEEtitleabstractindextext{\begin{abstract}Dynamic manipulation of the surface plasmon polariton (SPP) and
wave steering are important in plasmonic applications. In this work, we
excite a curved SPP in topological continua by applying a radial magnetic
bias. We believe that it is a new technique to create a unidirectional SPP
traveling along a curved trajectory. We also derive a Green's function model for radially-biased plasma, applicable to curved
SPPs. We compare the properties of unidirectional curved SPPs with the usual
case when an axial bias is applied.\end{abstract}

}

\maketitle

\section{INTRODUCTION}

\IEEEPARstart{D}{ifferent} techniques can be applied to dynamically manipulate the
propagation direction of the surface plasmon polariton (SPP). Directional
SPPs can be excited by engineering the design of SPP
launchers, for example designing metasurfaces \cite{lin2013polarization},
simple metallic gratings coated by nonlinear optical materials \cite%
{chen2013submicron}, asymmetric gratings, slits and resonators \cite%
{lopez2007efficient, liu2012compact}, grooves with different depth and width 
\cite{yao2015efficient}, and changing the incident wave polarization \cite%
{rodriguez2013near}; see Refs. \cite{ding2019review, wang2019dynamical} for
comprehensive reviews. In these cases, even though the directionality is
tunable, the excited SPPs still have a linear trajectory. However, they can
be effectively guided along a curvature by applying graded index (GRIN)
photonic crystals with a nonuniform refractive index \cite%
{zentgraf2011plasmonic} or patterned structures (e.g. \cite{cui2013thz}). In
addition, 2D materials such as graphene, whose optical properties are
electronically tunable, provide a good platform for directing SPPs along
even right-angled curvatures \cite{pakniyat2021reflectionless}. Nonetheless,
SPPs directed using these techniques are not inherently reflection-free. In
this regard, Airy SPP beams and hook SPPs are known as self-bending and
diffraction-free surface waves. They propagate along a parabolic trajectory.
Airy beams are generated by applying a spatial light modulator (SLM) or a
composite optical element with cubic phase. Illuminating Airy beams into a
simple grating or applying a metasurface providing the required cubic phase,
leads to excitation of Airy SPPs \cite{zhang2011plasmonic,
bleckmann2013manipulation}. Due to poor operation of SLM in the terahertz
frequency range, a more complex mechanism is required to excite THz Airy SPPs.
Surface plasmon polariton Bessel beams are another type of diffraction-free
surface waves that are generated by a similar mechanism as Airy SPP beams, but
they have a linear trajectory \cite{xiao2014dynamic, lin2012cosine}. Plasmonic
hook beams are newly-discovered curved SPPs, which are generated using a
simple asymmetric prism \cite{minin2018photonic, minin2021experimental}.
However, their curved trajectory exists only in the near-field.
Another possibility is an SPP vortex, which is an electromagnetic wave
carrying orbital angular momentum. It is excited using spiral slits \cite%
{zang2019manipulating} or nanoslits that provide the required
phase difference \cite{kim2010synthesis}.

In this work, we use the concept of topological insulators to
obtain a unidirectional SPP traveling in a circular path. We find
that by applying a radial magnetic field bias, SPPs that travel along a
curved trajectory are excited at the interface of the isotropic and radially
biased plasma media. The excited SPPs are unidirectional and
reflection-free. The surface waves are resistant to disorder because of
their one-way propagation properties, which results in longer propagation
even along, say, rough surfaces or surfaces with discontinuities. Their
properties are tunable by the magnetic field intensity as well as
frequency. The unidirectional curved SPP propagates on the surface of a homogeneous medium, and there is no
need to apply a grating or other structural pattern with narrow
bandwidth to steer SPPs in a circular path. As a result, better performance,
higher power transmission and wider bandwidth are achievable.

In continuous plasmonic materials such as metals and
semiconductors, a static magnetic field induces
a gyrotropic response and results in non-reciprocity due to broken time
reversal symmetry; the magnetized plasma is categorized as a
photonic topological insulator (PTI). One of the most 
 important aspects of
PTIs is their ability to support unidirectional SPPs with unique properties,
such as propagation in one direction, and protection from
back-scattering and diffraction upon encountering a discontinuity 
\cite{silveirinha2016bulk,shastri2021nonreciprocal}. They are also
characterized by an integer Chern invariant, indicating the number of
topologically protected surface modes. This number cannot change except when
the topology of the bulk bands is changed \cite%
{sil2015chern,tauber2020anomalous,gangaraj2017berry}.

The properties of unidirectional SPPs have been widely studied in 
systems biased by an in-plane axial bias \cite%
{gangaraj2020broadband,monticone2020truly,
pakniyat2020non,zhang2019backscattering}. In the well-known Voigt
configuration, the SPPs travel along a straight line perpendicular to the
in-plane axial magnetic bias vector at frequencies in the band gap above the
plasma frequency \cite{davoyan2013theory,gangaraj2016effects, tunable}.
However, in this work we realize that by applying a radial magnetic field,
similar propagation behavior is observed in that frequency regime, i.e. SPPs
tend to propagate perpendicular to the radial bias at the interface between gyrotropic and isotropic media. In fact, this new
configuration suggests the excitation of SPPs with circular trajectory due
to applying the radial bias. Hence, the SPP direction is steerable by
rotation of the magnetic bias direction. Using this technique, SPPs can be
effectively guided at right-angled bends. To analytically
investigate the properties of the unidirectional curved SPPs, we derive a
dyadic Green's function (GF) for a radially magnetized plasma. 

Dynamic manipulation of SPPs is of great interest. Like other types of curved SPPs, unidirectional curved SPPs can be used in applications such
as plasmonic tweezers, particle manipulation, bio-plasmonic systems,
switches and energy routing in plasmonic circuitry. Moreover, they can be
used in design of nonreciprocal devices, such as plasmonic circulators or in
generating hotspots \cite{zhang2021plasmonic,fang2015nanoplasmonic,
gangaraj2020broadband}.

In the following, we describe the curved topological SPPs and the
required conditions for their excitation. Then, we explain our Green's
function model and model, and provide a comparison with the numerical
results based on the finite element method using COMSOL. We discuss the
effect of different parameters on properties of the azimuthally propagating
SPPs. Finally, we propose an application for the curved SPPs.

\section{Curved Surface Plasmon Polaritons}

\begin{figure*}[tbp]
\includegraphics[width=2.0\columnwidth]{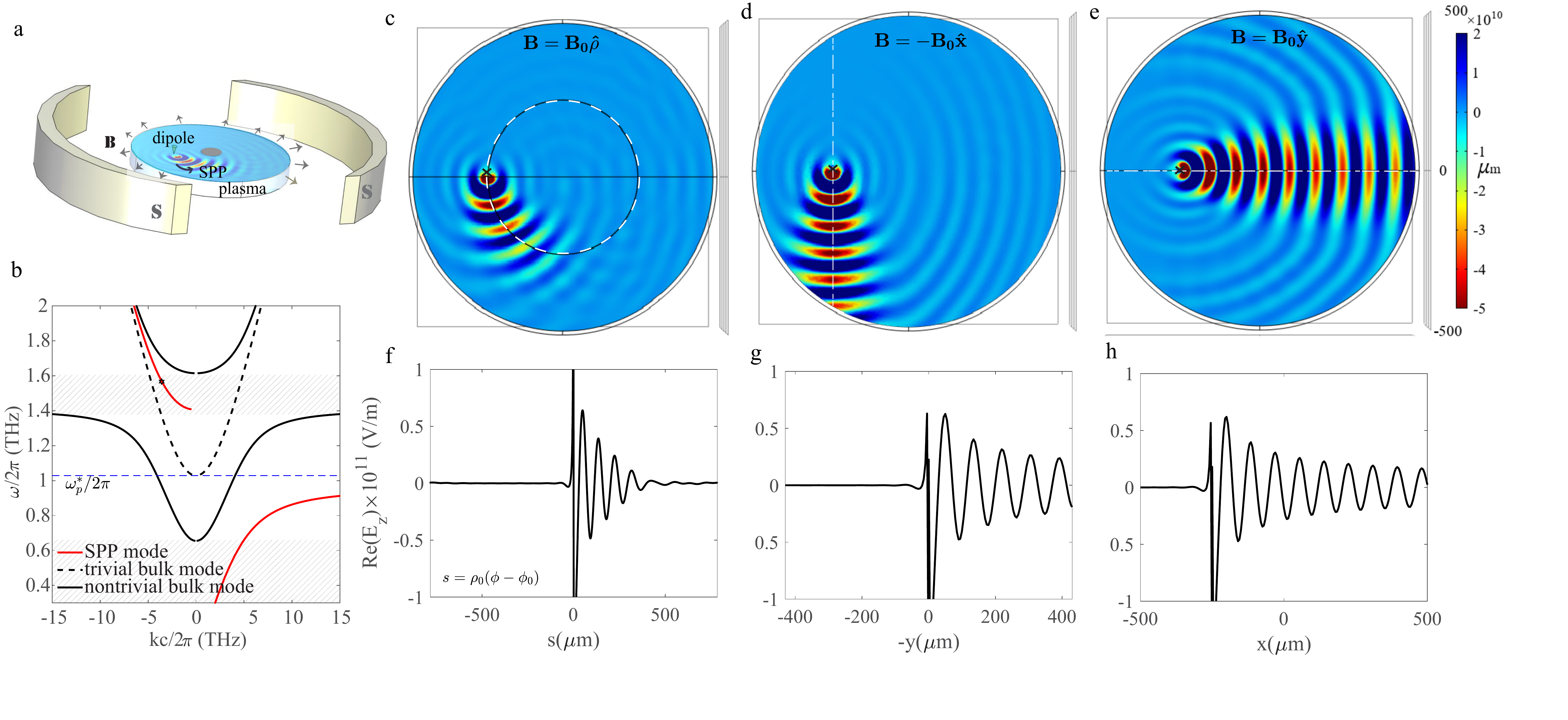}  
\caption{Unidirectional curved SPP vs. linear propagating SPP. (a) Geometry sketch (b) Bulk and
SPP dispersion diagrams. The gray regions denote the band gaps. The electric
field profile ($E_z$) of the SPP propagating at the interface of an
isotropic medium and a plasma region magnetized by (c) a static radial bias $%
\mathbf{B}=B_{0}\mathbf{\hat{\protect\rho}}$ that is centered on the origin,
(d) an axial bias $\mathbf{B}=-B_{0}\mathbf{\hat{x}}$, and (e) an axial bias 
$\mathbf{B}=B_{0}\mathbf{\hat{y}}$. The SPPs are excited by a point source
located at $(\protect\rho_{0},\protect\phi_{0},z)=(250\protect\mu\text{m},%
\protect\pi,0)$. (f,g,h) The electric field oscillation, respectively along
the circular, vertical and horizontal trajectories shown by white dashed
lines in (c,d,e). In (f), s is the arc length defined as $s=\protect\rho%
_{0}cos(\protect\phi-\protect\phi_{0})$ where $\protect\phi$ is the angle of
the observation point with respect to the x-axis and $\protect\rho_{0}$ is
the radius of the dashed circle. The magnetized plasma is characterized by (%
\protect\ref{epbc}) where $n_e=3.6\times 10^{21}\text{m}^{-3},
m^{\ast}=0.0175m_{0}, \protect\varepsilon_{\infty}=15.68, B_0=0.6\text{T}$,
given $\protect\omega_p^{\ast}=\protect\omega_p/\protect\sqrt{\protect%
\varepsilon_{\infty}}=2\protect\pi(1.03 \text{THz}), \protect\omega_{c}/%
\protect\omega_{p}^{\ast}=0.93$ and $\Gamma=0.00015\protect\omega_{p}$. The
top region is metal with dielectric constant of $\protect\varepsilon%
_r=-10^{4}$. The resonance frequency $f=1.567 \text{THz}$ is within the
upper band gap. }
\label{DISP}
\end{figure*}

Consider a plasma medium consisting of $n_{e}$ free electrons with
the effective mass of $m^{\ast }$ per volume, which is magnetized by a
static magnetic field bias $\mathbf{B}_{c}=B_{0}\mathbf{\hat{b}}_{c}$ where $%
B_{0}$ is the magnetic field strength and $\mathbf{\hat{b}}_{c}$ is a unit
vector along the direction of the magnetic field. In
general, the material is characterized by a dielectric tensor \cite%
{chen1983theory}

\begin{equation}
\mathbf{\bar{\varepsilon}=}\varepsilon _{t}(\mathbf{\bar{I}-\hat{b}}_{c}%
\mathbf{\hat{b}}_{c})+i\varepsilon _{\text{g}}(\mathbf{\hat{b}}_{c}\times 
\mathbf{\bar{I}})+\varepsilon _{\text{a}}\mathbf{\hat{b}}_{c}\mathbf{\hat{b}}%
_{c}  \label{epbc}
\end{equation}%
where the permittivity elements are defined using a Drude model as

\begin{gather}
\varepsilon _{t}=\varepsilon _{\infty }-\frac{\omega _{p}^{2}(1+i\Gamma
/\omega )}{(\omega +i\Gamma )^{2}-\omega _{c}^{2}},\text{ \ }\varepsilon
_{a}=\varepsilon _{\infty }-\frac{\omega _{p}^{2}}{\omega (\omega +i\Gamma )}%
,\text{ \ } \notag \\
\varepsilon _{g}=\frac{\omega _{c}\omega _{p}^{2}}{\omega \left[
\omega _{c}^{2}-(\omega +i\Gamma )^{2}\right] },  \label{e1}
\end{gather}%
assuming the time harmonic variation of $e^{-i\omega t}$; $\omega _{p}=$ $%
\sqrt{n_{e}q_{e}^{2}/(m^{\ast }\varepsilon _{0})}$, $\omega _{c}=$ $%
-q_{e}B_{0}/m^{\ast }$ and $\Gamma =-q_{e}/\mu m^{\ast }$ are plasma,
cyclotron, and collision frequencies, respectively, $q_{e}$ is the electron
charge, $\varepsilon _{\infty }$ is high-frequency dielectric constant, and $%
\mu $ is the carrier mobility. In this work, we apply a uniform radial bias, 
$\mathbf{B}_{c}=B_{0}\mathbf{\hat{\rho}}$ to magnetize the plasma region.
Figure \ref{DISP}a illustrates the geometry scheme of the system under
study. It includes a plasma slab under radial bias, covered by an isotropic
material. The radial bias can be practically implemented using concave permanent magnets. We assume that the plasma thickness is large,
then the system is two half-spaces of gyrotropic/isotropic media. The plasma
region is model by (\ref{epbc}), where $\mathbf{\hat{b}}_{c}=\mathbf{\hat{%
\rho}}$ and $\mathbf{\bar{I}}$ is a dyadic tensor in polar
coordinates with ($\mathbf{\hat{\rho},\hat{\phi},\hat{z}}$) unit
basis. Next, we study the properties of the bulk modes propagating inside
the radially magnetized plasma region. Then, we look for the SPPs excited at
the interface of the isotropic/radially magnetized plasma media.

A plane wave propagating in the gyrotropic medium with the wave vector $%
\mathbf{k}=k_{\rho }\mathbf{\hat{\rho}}^{\prime }\mathbf{+}k_{z}\mathbf{%
\mathbf{\hat{z}}}$ satisfies the wave equation $\mathbf{k}\times \left( 
\mathbf{k\times E}\right) +k_{0}^{2}\mathbf{\bar{\varepsilon}}_{r}\cdot 
\mathbf{E=0}$. The non-zero solution of $\mathbf{E}$ exists only if 
$\left\vert k_{0}^{2}\mathbf{\bar{\varepsilon}}_{r}-k^{2}\mathbf{\bar{I}}+%
\mathbf{kk}\right\vert =0$. This determinant is the dispersion equation of
the bulk modes propagating with an arbitrary direction in a gyrotropic
medium. Consider an orthogonal coordinate system, having a unit vector along
the magnetic bias as $\left\{ \mathbf{\hat{k}}_{t},\mathbf{\hat{\rho},\hat{k}%
}_{t}\times \mathbf{\hat{\rho}}\right\} $. The wave vector in this
coordinate is rewritten as $\mathbf{k=k}_{t}+q_{\rho }\mathbf{\hat{\rho}}$
with $\mathbf{k}_{t}=q_{\varphi }\mathbf{\hat{\phi}+}k_{z}\mathbf{%
\hat{z}}$, where $q_{\rho }=k_{\rho }\cos (\phi _{\mathbf{k}}-\phi _{\mathbf{%
b}})$ and $q_{\varphi }=k_{\rho }\sin (\phi _{\mathbf{k}}-\phi _{\mathbf{b}})
$; $\phi _{\mathbf{k}}$ and $\phi _{\mathbf{b}}$ are the angle of the wave
and bias vectors with respect to the $x$-axis, respectively. By plugging $%
\mathbf{\bar{\varepsilon}}$ and $\mathbf{k}$ into the above determinant, we
derive%
\begin{equation}
k_{t}^{2}=\frac{1}{2\varepsilon _{t}}\left[ -\kappa \pm \sqrt{\kappa
^{2}-4\varepsilon _{t}\nu }\right] \text{,}  \label{kt}
\end{equation}%
where $\kappa =q_{\rho }^{2}(\varepsilon _{t}+\varepsilon
_{a})+k_{0}^{2}\left( \varepsilon _{g}^{2}-\varepsilon _{t}(\varepsilon
_{t}+\varepsilon _{a}\right) )$, $\nu =\varepsilon _{a}\left( q_{\rho
}^{2}-k_{0}^{2}\varepsilon _{t}\right) ^{2}-\varepsilon _{a}\varepsilon
_{g}^{2}k_{0}^{4}$, and $k_{t}^{2}=q_{\varphi }^{2}+k_{z}^{2}$. We look for
the bulk modes propagating perpendicular to the bias. Thus, we set $q_{\rho
}=0$ in (\ref{kt}) and determine two equations as $k_{1}^{2}=\varepsilon _{%
\text{eff}}k_{0}^{2}$ and $k_{2}^{2}=\varepsilon _{\text{a}}k_{0}^{2}$ where 
$\varepsilon _{\text{eff}}=(\varepsilon _{t}^{2}-\varepsilon
_{g}^{2})/\varepsilon _{t}$, $\ k_{0}$ is the free space wave number, and $%
k_{j}^{2}=q_{\varphi }^{2}\mathbf{+}k_{zj}^{2}$, $j\in \left\{ 1,2\right\} $%
. These equations characterize the nontrivial TM modes with $E_{\varphi
}, E_{z}, H_{\rho }$ (no electric field component along the bias vector) and
trivial TE modes with $H_{\varphi }, H_{z}, E_{\rho }$ (no magnetic
field component along the bias), respectively. The
nontrivial modes are dependent to the magnetic bias, unlike the trivial
modes. Note that in a cylindrical rode pure TE and pure TM modes exist only when the field configurations are symmetric and independent of $\phi$. Here, nontrivial TM and trivial TE modes have phase variation of $\exp (im\varphi )$. Therefore, they cannot be pure TE and TM modes; they are hybrid modes. A wave with $q_{\rho }=0$ is a traveling wave on a cylindrical
shell, which can be decomposed into nontrivial TM\ and trivial TE modes in a radially magnetized system. It has a vortex-like behavior
and its phase varies as $\exp (im\varphi )$. An electromagnetic vortex is a
differentiated plane wave which can be generated by three homogeneous plane
wave interference \cite{masajada2001optical,hannay2015differentiated}.

Next, by enforcing continuity of the tangential components of
the electric and magnetic fields of these particular bulk modes at the
interface, we derive the SPP dispersion equation as 
\begin{equation}
\frac{\sqrt{k_{s}^{2}-k_{0}^{2}\varepsilon _{r}}}{\varepsilon _{r}}+\frac{%
\sqrt{k_{s}^{2}-k_{0}^{2}\varepsilon _{\text{eff}}}}{\varepsilon _{\text{eff}%
}}=\frac{\varepsilon _{g}k_{s}}{\varepsilon _{t}\varepsilon _{\text{eff}}},
\label{SPP}
\end{equation}%
where $k_{s}=q_{\varphi }$ is the propagation constant of the surface wave
and $\varepsilon _{r}$ is the effective permittivity of the isotropic
region. This dispersion relation is the same as for axial bias in the Voight configuration.\ Figure \ref%
{DISP}b shows the dispersion diagrams of the nontrivial TM, trivial TE bulk
modes, and the SPP modes. The shaded gray regions indicate
bandgaps between the nontrivial bulk bands. Like usual
topological plasma systems when an axial bias is applied, the SPPs crossing
the nontrivial bandgaps are potentially topological. Their frequency
response is asymmetric.

Then, We have simulate the system under study using COMSOL\ Multiphysics. The
plasma region is characterized by parameters presented in \cite%
{pakniyat2020indium,liang2021temperature}, and provided in the caption of Fig. 1, related to an undoped InSb
crystal at moderate temperatures. The SPPs are excited by a point source
located at the interface of the gyrotropic/isotropic media, operating at a
frequency within the upper nontrivial bandgap. The electric field profile at
the interface is shown in Fig. \ref{DISP}c. It shows the SPPs propagating
counter clock-wise (CCW) on a circular path about the origin. There is no
propagation in the opposite direction due to the unidirectional nature of
the wave. So, the excited SPPs have a circular trajectory rather than a
linear trajectory as a result of applying the radial bias.

For comparison, we obtain the field profile of the SPPs when the axial
biases $\mathbf{\hat{b}}_{c}=-\mathbf{\hat{x}}$ and $\mathbf{\hat{b}}_{c}=%
\mathbf{\hat{y}}$ are applied (the usual cases). The results are shown in
Figs. \ref{DISP}d and \ref{DISP}e. Here, the unidirectional SPPs have
linear propagation. They are characterized by the same dispersion
equation as Eq. \ref{SPP}, but with surface momentum $k_{s}=-q_{y}$ and $k_{s}=q_{x}$, respectively. Comparing Fig. \ref{DISP}c
with Fig. \ref{DISP}d and Fig. \ref{DISP}e, the deviation of the SPPs from a
straight line to a circular path is evident. In all cases, the SPPs tend to
propagate perpendicular to the static bias. For that reason, in the radial
bias system the surface plasmons gain orbital angular momentum and
form curved SPPs.

The line graphs in Fig. \ref{DISP}f,g,h indicate the electric field
oscillation along the circular, vertical and horizontal straight line traces
shown by white dashed lines in the field profile plots. According to the
period of the oscillation, the SPP wavelength for radial and axial bias
cases are almost equal ($\lambda _{SPP}\simeq 84\mu m$). The obtained
wavelength is consistent with the estimated value obtained from the
dispersion diagram ($\lambda _{SPP}=2\pi /\text{Re}(k_{s})$). The SPPs have
similar propagation properties, however, the decay rate of the curved SPP is much higher. We find that the curved SPPs are leaky modes,
while the linear SPPs in the axially biased systems are confined propagating
modes. The difference in results is due to the hybrid
nature of the nontrivial and trivial bulk modes in the radially biased
system. In fact, the curved SPPs excited at a resonance frequency within the
upper nontrivial bandgap can be coupled to the trivial TE cylindrical bulk
modes. This does not occur in an axially biased system, because in that case
the trivial TE and nontrivial TM modes are orthogonal modes, and hence, the
TE modes do not contribute to the excitation of the TM SPP. Consequently,
the TM SPPs are confined modes at frequencies within the nontrivial bandgaps
and their energy does not couple to the trivial TE\ mode.
\section{Dyadic Green's Function for a radially magnetized plasma}

\begin{figure*}[tbp]
\begin{center}
\includegraphics[width=1.7\columnwidth]{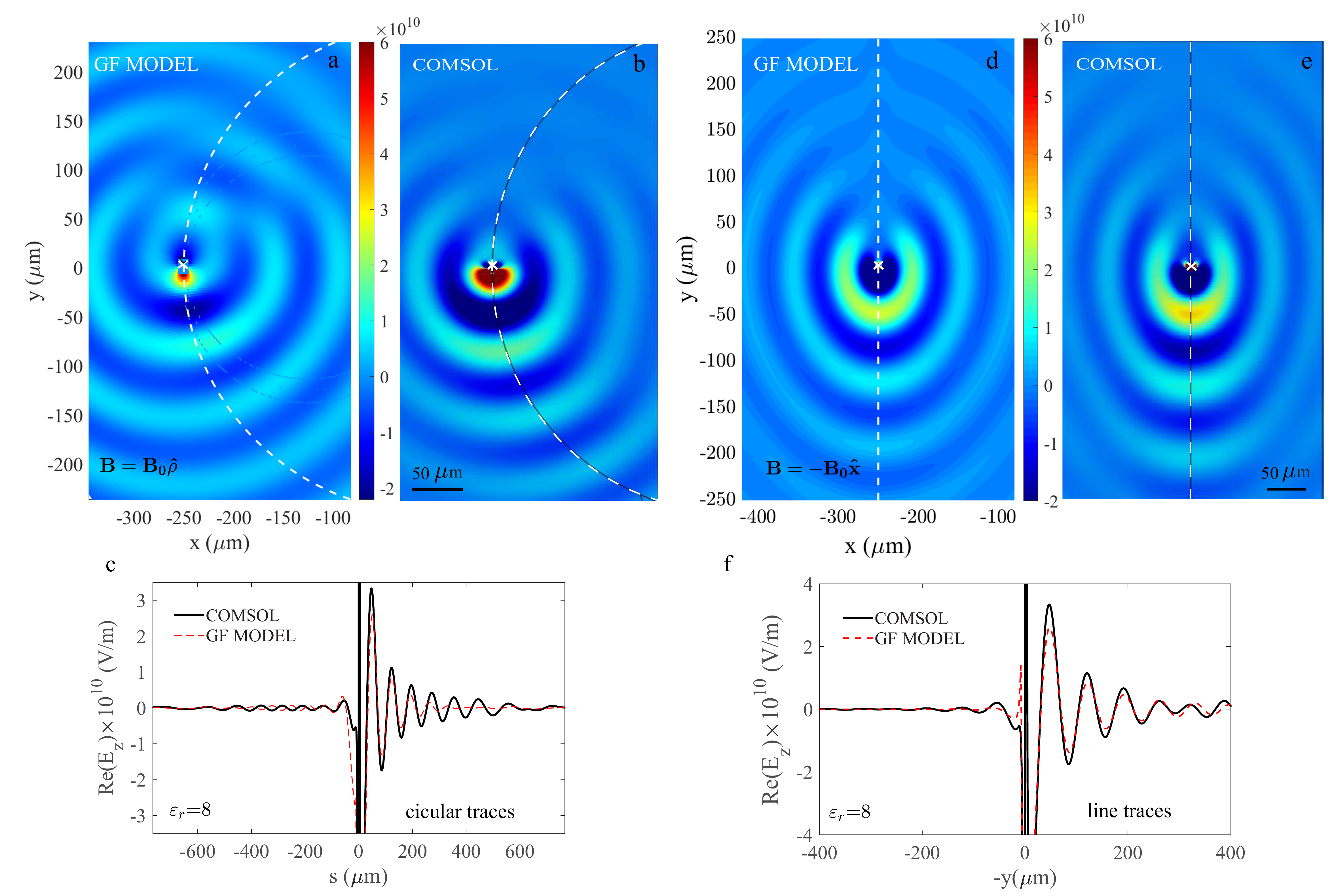}  
\caption{Electric field distribution ($E_{z}$) of the curved SPPs excited at
the interface of dielectric/ radially magnetized plasma ($\mathbf{B}=B_{0}%
\mathbf{\hat{\protect\rho}}$) media using (a) Green's function model (%
\protect\ref{Ezs}) and (b) a full-wave COMSOL simulation. The radial bias is
centered on the origin and the dipole is located at $(250\protect\mu\text{m},%
\protect\pi,0)$. The observation points are on a plane with distance $z=0.003%
\protect\lambda_{p}$ above the interface ($\protect\lambda_{p}=2\protect\pi/%
\protect\omega_{p}$). (c) The electric field oscillation along the circular
trajectories (white dashed semicircles). The electric field profile of the
linear propagating SPPs excited at the interface of the
dielectric/axially biased plasma ($\mathbf{B}=-B_{0}\mathbf{\hat{x}}$) using
(d) GF model (e) COMSOL simulation. (f) The extracted data from dashed line
trajectories. The magnetized plasma is characterized by (\protect\ref{epbc})
where $n_e=3.6\times 10^{21}\text{m}^{-3}, m^{\ast}=0.0175m_{0}, \protect%
\varepsilon_{\infty}=15.68, B_0=0.6\text{T}$ and $\Gamma=0.00015\protect%
\omega_{p}$. The dielectric constant of top region is $\protect\varepsilon%
_r=8$. The resonance frequency is $f=1.567 \text{THz}$. }
\label{GFCO}
\end{center}
\end{figure*}

Here, we analytically obtain the electric field of the curved SPPs in a
radially magnetized system. They are excited by a point source at the
interface of two half-space media where the $z<0$ region is filled
by a radially magnetized plasma and the $z>0$ region is an isotropic
material. The radial bias is centered on the origin and the dipole is
located inside the isotropic region at $\mathbf{r}_{0}=(\rho _{0},\phi
_{0},d)$. By doing a Green's function (GF) analysis in a polar coordinate system,
the tangential and normal components of the scattered field in the isotropic
region at $\mathbf{r}=(\rho ,\phi _{\mathbf{r}},z)$ due to a vertical dipole
source with moment of $\mathbf{p=}\gamma \mathbf{\hat{z}}$ are governed by 
\begin{equation}
\mathbf{E}_{||}^{s}(\mathbf{r})=\frac{1}{(2\pi )^{2}}\int_{0}^{\infty
}\int_{0}^{2\pi }\mathbf{\bar{R}(q)\cdot }i\mathbf{q}\frac{\gamma e^{-\gamma
_{0}(z+d)}}{2\varepsilon _{r}\varepsilon _{0}}e^{i\mathbf{q}\cdot \left( 
\mathbf{r}-\mathbf{r}_{0}\right) }qd\phi _{\mathbf{q}}dq  \label{Es1}
\end{equation}%
and 
\begin{equation}
E_{z}^{s}(\mathbf{r})=\frac{-1}{(2\pi )^{2}}\int_{0}^{\infty }\int_{0}^{2\pi
}C^{r}(\mathbf{q})\frac{%
\gamma e^{-\gamma _{0}(z+d)}}{2\gamma _{0}\varepsilon _{r}\varepsilon _{0}}e^{i\mathbf{q}\cdot \left( \mathbf{r}-%
\mathbf{r}_{0}\right) }qd\phi _{\mathbf{q}}dq  \label{Ezs1}
\end{equation}%
where $C^{r}(\mathbf{q})=\mathbf{q}\cdot \mathbf{\bar{R}(q)\cdot q}$ and $%
\mathbf{\bar{R}(q)}$ is a 2x2 reflection coefficient. The GF derivation
details and quantities are defined in the
appendix. Using these 2D Sommerfeld integrals, the field computation is very
time consuming and it does not converge well. To solve this problem, we use
a saddle point approximation and simplify (\ref{Ezs1}) to a 1D integral as
defined in (\ref{SD1}). Then, using this Green's function model, we generate
the field density profile shown in Fig. \ref{GFCO}a by computing the
electric field of the observation points on a plane above the interface with
local position ($\rho ,\phi _{\mathbf{r}},z=0.003\lambda _{p}$), where $%
\lambda _{p}=2\pi /\omega _{p}$. The top region is a dielectric with $%
\varepsilon _{r}=8$. The plot shows one-way SPPs with CCW propagation on a
circular path. For comparison, we generated Fig. \ref{GFCO}b based on a
numerical computation using COMSOL. The GF result is consistent with the
numerical result. Then, the data are extracted from the circular traces
shown by white dashed lines to generate the line graph in Fig. \ref{GFCO}c.
As shown, the results arising from the GF model are very close to the COMSOL
results, which validates the accuracy of our GF model for radially biased
system.

We also develop the GF model presented in \cite{silveirinha2018fluctuation,
silveirinha2014optical} for an axial bias along the $-x$ direction. For this
case, the SPPs are propagating along a straight line. Figure \ref{GFCO}d and %
\ref{GFCO}e demonstrate the electric field density and Fig. \ref{GFCO}g
shows the SPP oscillation along the dashed line trajectories using axial GF
model and COMSOL simulation.

\begin{figure*}[tbp]
\begin{center}
\includegraphics[width=2\columnwidth]{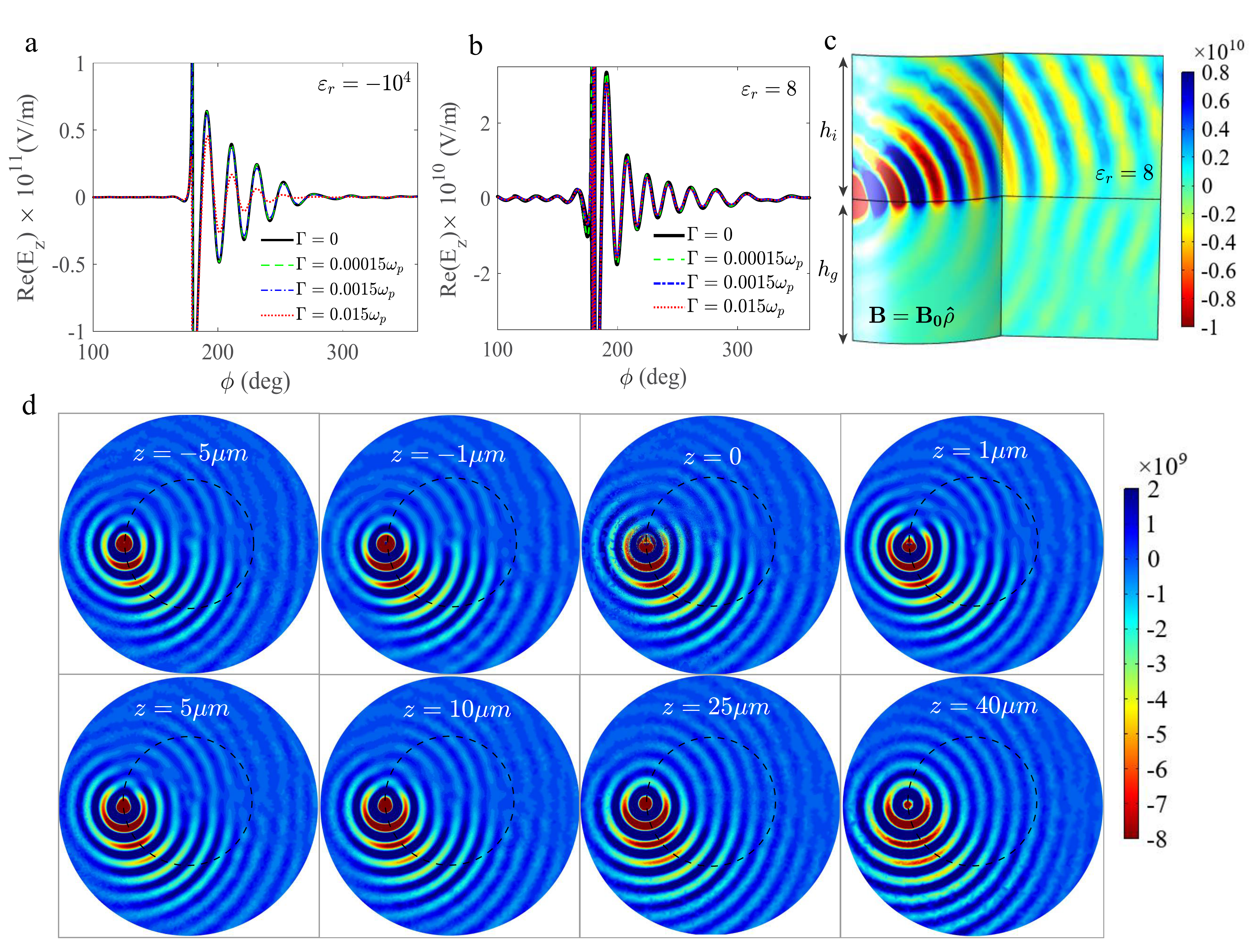}  
\caption{(a,b) Unidirectional curved SPPs by considering different amounts
of dissipation (metal or dielectric on top) (c) A vertical cross-section
view of the two-layer magnetized system ($h_i=h_g=200\protect\mu\text{m}$).
Energy leakage occurs in the dielectric region after several
wavelengths of curved SPP propagation (d) Electric field density of the
curved SPPs on the surfaces parallel to the interface located at different
distances (z) from the interface $z=0$ ($\protect\varepsilon_r=8$). $%
n_e=3.6\times 10^{21}\text{m}^{-3}, m^{\ast}=0.0175m_{0}, \protect\varepsilon%
_{\infty}=15.68, B_0=0.6\text{T}$, $f=1.567 \text{THz}$.}
\label{prop}
\end{center}
\end{figure*}

Figure \ref{prop}a,b shows the unidirectional curved SPP oscillation along
the circular path in the radially magnetized system by considering different
amounts of dissipation when the top region is metal or dielectric. As shown, by reducing the loss, the magnitude
increases and the curved SPPs propagate longer, as expected. However, even in
a loss-less system the SPPs do not rotate on a full circle. They stop their
orbital propagation on the surface after several SPP wavelengths of propagation
and they radiate to the plasma or dielectric region. The leakage to the dielectric is
illustrated in Fig. \ref{prop}c, showing a vertical cross section of the
system (including the dielectric and the radially magnetized plasma); a cut
cylinder that is intersected by a plane. The curved SPP is leaky for
this operating frequency, as discussed in Section I.
In addition, for the case of dielectric on top, the mode lies within the light cone of the dielectric region. 
Figure \ref{prop}d shows the electric field profile on surfaces parallel
to the interface, located at different heights below and above the
interface. In the plasma region and close to the surface, the SPPs spiral on
a circle centered at the origin. In 
the dielectric region, they remain on this path at distances close
to the interface. Moving farther vertically from the interface, SPPs spiral
out of the circle. We also observed that SPPs are more confined to the
surface when the top layer is a metal.

\section{An application for curved SPPs}

\begin{figure*}[tbp]
\begin{center}
\includegraphics[width=1.5\columnwidth]{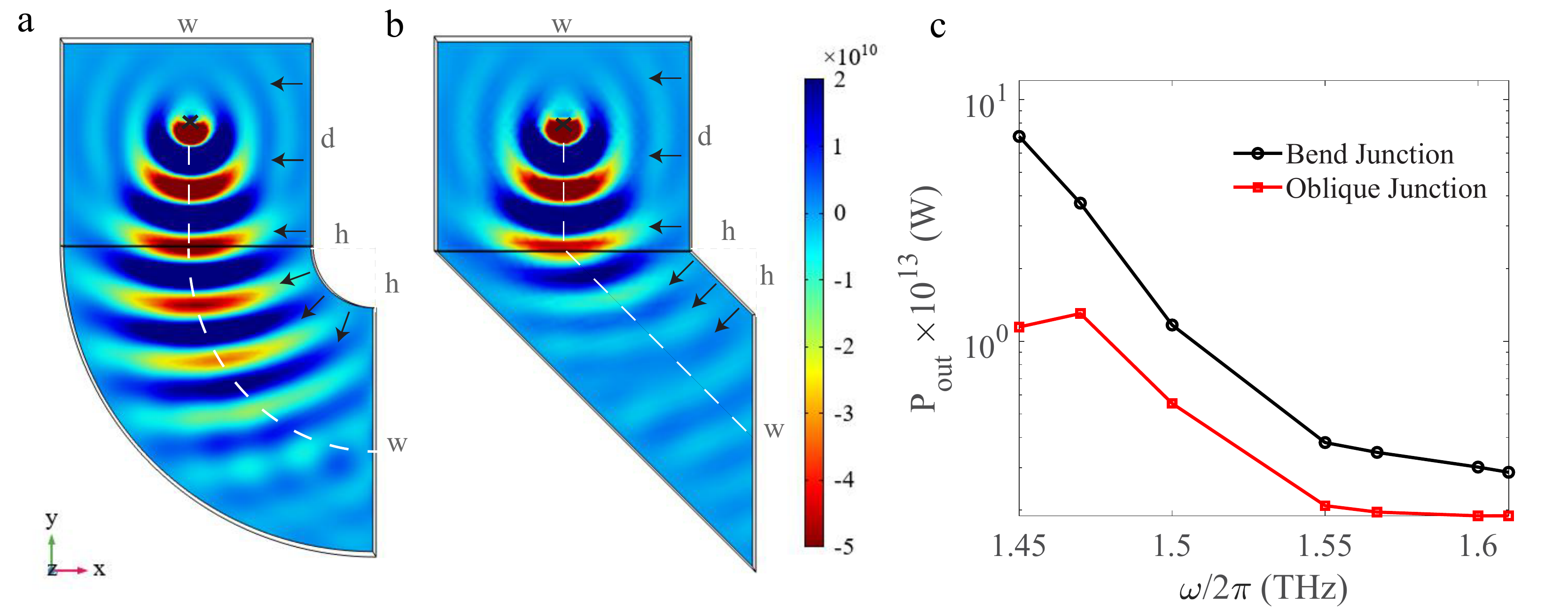}  
\caption{Electric field profile of the SPP propagating at the interface of
metal/magnetized plasma, passing through (a) a $90^\circ$ circular bend
junction under radial bias vs. (b) an oblique junction under $5\protect\pi/4$
axial bias. (c) Output power vs. frequency. $n_e=3.6\times 10^{21}\text{m}%
^{-3}, m^{\ast}=0.0175m_{0}, \protect\varepsilon_{\infty}=15.68, B_0=0.6%
\text{T}$, $\Gamma=0.00015\protect\omega_{p}$, $\protect\varepsilon_r=-10^{4}
$, $f=1.567 \text{THz}, w=4h=360\protect\mu\text{m}, d=300\protect\mu\text{m}%
.$ }
\label{junc}
\end{center}
\end{figure*}

Waveguide bends connecting two straight waveguides are important components
in plasmonic integrated circuits. Using unidirectional curved SPPs, a bent
waveguide with minimal bending loss can be designed. We propose that a
90-degree circular bend magnetized by a radial bias can be used as a
nonreciprocal plasmonic junction. As shown in Fig. \ref{junc}a, the excited
unidirectional SPPs steer from a straight line to a circular path through
the 90 degree bend, resulting in reduction of the radiation loss due to the
curvature of the waveguide junction. Black arrows indicate the magnetic bias
vectors applied in each segment. It forms an optical
nonreciprocal plasmonic junction, which allows power transmission only in
one direction.
To provide a comparison to the axial-bias case, in Fig. \ref{junc}b, two
straight waveguides are connected by an oblique junction magnetized by an
axial bias with angle of $5\pi /4$ radian. When unidirectional SPPs reach the input port of the oblique junction, they change direction and align themselves along a line perpendicular to the bias. That is because the unidirectional SPPs inherently tend to
propagate perpendicular to the magnetic bias at frequencies within the nontrivial bandgap.

The surface power that flows through these two junctions are computed at the
output ports for different operating frequencies within the upper bandgap
and shown in Fig. \ref{junc}c. The power is transmitted through the radially
magnetized circular bend more than two times higher than the power
transmitted through the oblique junction. In addition, the power transmission is
significantly higher than an unbiased circular junction. In the circular
bend with radial bias, the energy routing only occurs in one
direction. By reversing the magnetic field direction, the energy is routed in the opposite direction.

\section{Conclusion}

In conclusion, we obtained unidirectional curved SPPs by applying an
in-plane radial magnetic bias in topological continua. In a magnetized
system, the unidirectional SPP trajectories are steerable by the
magnetic bias direction. We derived a Green's function model for a radially
magnetized system. The properties of unidirectional curved SPPs were
compared to the linear SPP. Using unidirectional curved SPPs, a bent
waveguide with minimal bending loss and nonreciprocal features was proposed
for plasmonic integrated circuits.

\section{Appendix}

\begin{figure}[tbp]
\includegraphics[width=1\columnwidth]{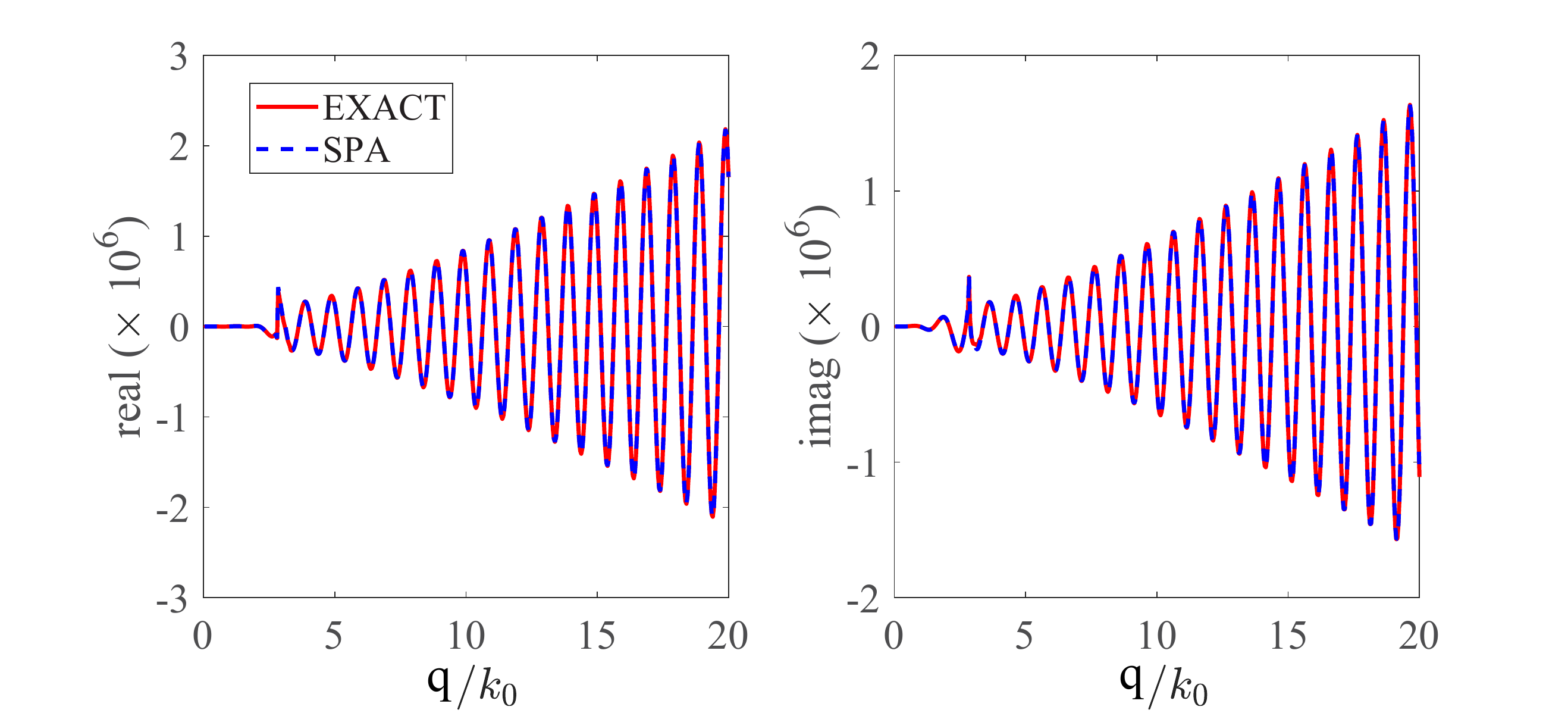}  
\caption{Saddle point approximation vs. exact solution (the integrand of (%
\protect\ref{SD1}) and (\protect\ref{Ezs}), respectively, over $q$
variation) for $f=1.567\text{THz}$ at ($\protect\rho =250\protect\mu \text{m}%
,\protect\phi =5\protect\pi /4,z=0.003\protect\lambda _{p}$). $%
n_{e}=3.6\times 10^{21}\text{m}^{-3},m^{\ast }=0.0175m_{0},\protect%
\varepsilon _{\infty }=15.68,B_{0}=0.6\text{T}$, and $\protect\varepsilon %
_{r}=8$.}
\label{SPA}
\end{figure}

Here, we obtain the scattered field in a radially biased system based on a
Green's function analysis in polar coordinates. Consider two half-space
isotropic/magnetized plasma media having an interface at $z=0$. An electric
source with dipole moment of $\mathbf{p=}\gamma \mathbf{\hat{z}}$ is located
at $\mathbf{r}_{0}=(\rho _{0},\phi _{0},d)$ in the isotropic region. The
primary electric and magnetic fields are $\mathbf{E}^{p}(\mathbf{r}%
)=(\varepsilon _{r}k_{0}^{2}\mathbf{I+\nabla \nabla })\cdot \mathbf{\pi }^{p}
$ and $\mathbf{H}^{p}(\mathbf{r})=i\omega \varepsilon _{0}\varepsilon _{r}%
\mathbf{\nabla }\times \mathbf{\pi }^{p}$, where the Hertzian potential $%
\mathbf{\pi }^{p}$ is given by $\mathbf{\pi }^{p}(\mathbf{r})=g^{p}(\mathbf{%
r,r}_{0})\mathbf{p/}\varepsilon _{r}\varepsilon _{0}$. The primary Green's
function is%
\begin{eqnarray}
g^{p}(\mathbf{r,r}_{0}) &=&\frac{e^{-i\sqrt{\varepsilon _{r}}k_{0}\left\vert 
\mathbf{r-r}_{0}\right\vert }}{4\pi \left\vert \mathbf{r-r}_{0}\right\vert }
\label{ed} \\
&=&\frac{1}{(2\pi )^{2}}\int_{0}^{\infty }\int_{0}^{2\pi }\frac{e^{-\gamma
_{0}\left\vert z-d\right\vert }}{2\gamma _{0}}e^{i\mathbf{q}\cdot \left( 
\mathbf{r}-\mathbf{r}_{0}\right) }qd\phi _{\mathbf{q}}dq \notag
\end{eqnarray}%
where $\mathbf{q}\cdot \mathbf{r}=q\rho \cos (\phi _{\mathbf{q}}-\phi _{%
\mathbf{r}})$ and $\mathbf{q}\cdot \mathbf{r}_{0}=q\rho _{0}\cos (\phi _{%
\mathbf{q}}-\phi _{0})$ with $\left\{ \phi _{\mathbf{q}},\phi _{\mathbf{r}%
},\phi _{0}\right\} $ denoting the angles $\mathbf{q}$, $\mathbf{r}$, and $%
\mathbf{r}_{0}$ make with the Cartesian unit vector $\mathbf{\hat{x}}$. The
exponential factor is $\gamma _{0}=\sqrt{q^{2}-\varepsilon _{r}k_{0}^{2}}$,
where $\varepsilon _{r}$ is the dielectric constant of the isotropic region (%
$z>0$). Using the Fourier transform pairs and nabla relations in a
polar coordinate, we have%
\begin{eqnarray}
\mathbf{E}^{p}(\mathbf{r}) &=&\frac{\gamma g^{p}(\mathbf{r,r}_{0})}{\varepsilon _{r}\varepsilon _{0}%
}\left[ \mathbf{\nabla }_{t}\partial _{z}+\left( \frac{\partial ^{2}}{%
\partial z^{2}}+\varepsilon _{r}k_{0}^{2}\right) \mathbf{\mathbf{\hat{z}}}%
\right] 
\end{eqnarray}
\begin{eqnarray}
\mathbf{E}^{p}(z,\mathbf{q}) &=&\mathbf{E}_{||}^{p}+E_{z}^{p}\mathbf{\mathbf{%
\hat{z}}} \notag \\ &=& \frac{\gamma g^{p}(z\mathbf{,q})}{\varepsilon _{r}\varepsilon _{0}}\left[ i\mathbf{q}%
\partial _{z}+\left( \frac{\partial ^{2}}{\partial z^{2}}+\varepsilon
_{r}k_{0}^{2}\right) \mathbf{\mathbf{\hat{z}}}\right] 
\label{ept}
\end{eqnarray}%
where $g^{p}(z\mathbf{,q})=e^{-i\mathbf{q}\cdot \mathbf{r}_{0}}e^{-\gamma
_{0}\left\vert z-d\right\vert }/2\gamma _{0}$. The total field in the
isotropic region is a superposition of the primary and scattered
field, $\mathbf{E}^{(1)}(\mathbf{r})=\mathbf{E}^{p}(\mathbf{r})+\mathbf{E}%
^{s}(\mathbf{r})$. Let $\mathbf{\bar{R}}$ be a reflection tensor such that
the tangential components of the scattered field at the interface are
related to the tangential primary field as 
\begin{equation}
\mathbf{E}_{||}^{s}(z\mathbf{,q})=\mathbf{E}_{||}^{s}(\mathbf{q})e^{-\gamma
_{0}z}=\mathbf{\bar{R}(\mathbf{q})\cdot E}_{||}^{p}(0\mathbf{,q})e^{-\gamma
_{0}z}.
\end{equation}%
Substitution of (\ref{ept}) gives%
\begin{equation}
\mathbf{E}_{||}^{s}(z\mathbf{,q})=\mathbf{\bar{R}(q)}\cdot i\mathbf{q}\frac{%
\gamma }{2\varepsilon _{r}\varepsilon _{0}}e^{-\gamma _{0}(z+d)}e^{-i\mathbf{%
q}\cdot \mathbf{r}_{0}}.  \label{Esp3}
\end{equation}%
According to the Gauss's law for the scattered field $\mathbf{\nabla }\cdot 
\mathbf{E}^{s}=0$, the $z$ component of the scattered field is $E_{z}^{s}(%
\mathbf{r})=-\int \mathbf{\nabla }_{t}\cdot \mathbf{E}_{||}^{s}(\mathbf{r})dz
$. Using (\ref{Esp3}) we have,%
\begin{eqnarray}
E_{z}^{s}(z\mathbf{,q}) &=&-i\mathbf{q}\cdot \mathbf{E}_{||}^{s}(\mathbf{q}%
)\int e^{-\gamma _{0}z}dz \notag \\
&=&\frac{-\gamma }{2\varepsilon _{r}\varepsilon _{0}\gamma _{0}}\mathbf{q}%
\cdot \mathbf{\bar{R}(q)\cdot q}e^{-\gamma _{0}(z+d)}e^{-i\mathbf{q}\cdot 
\mathbf{r}_{0}}.
\end{eqnarray}%
Finally, by taking the spatial Fourier transform we have%
\begin{equation}
\mathbf{E}_{||}^{s}(\mathbf{r})=\frac{1}{(2\pi )^{2}}\int_{0}^{\infty
}\int_{0}^{2\pi }\mathbf{\bar{R}(q)}\cdot i\mathbf{q}\frac{\gamma e^{-\gamma
_{0}(z+d)}}{2\varepsilon _{r}\varepsilon _{0}}e^{i\mathbf{q}\cdot \left( 
\mathbf{r}-\mathbf{r}_{0}\right) }qd\phi _{\mathbf{q}}dq
\end{equation}%
and%
\begin{equation}
E_{z}^{s}(\mathbf{r})=\frac{1}{(2\pi )^{2}}\int_{0}^{\infty }\int_{0}^{2\pi
}C^{r}(\mathbf{q})\frac{-\gamma }{\varepsilon _{r}\varepsilon _{0}}\frac{%
e^{-\gamma _{0}(z+d)}}{2\gamma _{0}}e^{i\mathbf{q}\cdot \left( \mathbf{r}-%
\mathbf{r}_{0}\right) }qd\phi _{\mathbf{q}}dq  \label{Ezs}
\end{equation}%
where $C^{r}(\mathbf{q})=\mathbf{q}\cdot \mathbf{\bar{R}(q)\cdot q}$. Using
the saddle point approximation, the last 2D Sommerfeld integral can be
simplified as%
\begin{equation}
E_{z}^{s}(\mathbf{r})=\int_{0}^{\infty }I(q)\frac{-\gamma e^{-\gamma
_{0}(z+d)}}{2\varepsilon _{r}\varepsilon _{0}\gamma _{0}}qdq  \label{SD1}
\end{equation}%
where%
\begin{eqnarray}
I(q)=\frac{1}{(2\pi )^{2}}\sqrt{\frac{\pi }{2q\left\vert g(\phi _{\mathbf{s}%
})\right\vert }}\left[ C^{r}(q,\phi _{\mathbf{s}})e^{iqg(\phi _{\mathbf{s}%
})}e^{i\pi \varsigma /4} \right. \notag \\ \left. +C^{r}(q,\phi _{\mathbf{s}}-\pi )e^{-iqg(\phi _{%
\mathbf{s}})}e^{-i\pi \varsigma /4}\right]   \label{SD2}
\end{eqnarray}%
with $g(\phi _{\mathbf{s}})=\rho \cos (\phi _{\mathbf{s}}-\phi _{\mathbf{r}%
})-\rho _{0}\cos (\phi _{\mathbf{s}}-\phi _{0})$, $\varsigma $ is the sign
of $g"(\phi _{\mathbf{s}})$, and the saddle point is%
\begin{equation}
\phi _{\mathbf{s}}=\tan ^{-1}\left( \frac{\rho _{0}\sin (\phi _{0})-\rho
\sin (\phi _{\mathbf{r}})}{\rho _{0}\cos (\phi _{0})-\rho \cos (\phi _{%
\mathbf{r}})}\right)   \label{SD3}
\end{equation}%
where $\phi _{\mathbf{s}}\in (0,\pi )$. Figure \ref{SPA} shows that the
approximated relation is matched with the exact solution. 

\subsection{Reflection Tensor in a radially biased system}

A plane wave in a gyrotropic medium satisfies the wave equation $\mathbf{k}%
\times \left( \mathbf{k\times E}\right) +k_{0}^{2}\mathbf{\bar{\varepsilon}}%
_{r}\cdot \mathbf{E=0}$. Consider a coordinate system, having a
unit vector along the magnetic bias $\left\{ \mathbf{\hat{k}}_{t},\mathbf{%
\hat{\rho},\hat{k}}_{t}\times \mathbf{\hat{\rho}}\right\} ,$ where $\mathbf{%
k=k}_{t}+q_{\rho }\mathbf{\hat{\rho}}$ with $\mathbf{k}%
_{t}=q_{\varphi }\mathbf{\hat{\phi}+}k_{z}\mathbf{\hat{z}}$. We define $%
q_{\rho }\equiv k_{\rho }\cos (\phi _{\mathbf{k}}-\phi _{b})$ and $%
q_{\varphi }\equiv k_{\rho }\sin \phi _{\mathbf{k}}-\phi _{b})$ where $%
q_{\rho },q_{\varphi }\in \left[ -\infty ,\infty \right] $. Note that the
permittivity tensor is given in ($\mathbf{\hat{\rho},\hat{\phi},\hat{z}}$)
polar basis which is related to ($\mathbf{\hat{\rho}%
{\acute{}}%
,\hat{\phi}%
{\acute{}}%
,\hat{z}}$) basis by projection relations of $\mathbf{\hat{\rho}%
{\acute{}}%
\cdot \hat{\rho}}\mathbf{=}\cos (\phi _{\mathbf{k}}-\phi _{b})$ and $\mathbf{%
\hat{\rho}%
{\acute{}}%
\cdot \hat{\phi}}\mathbf{=}\sin (\phi _{\mathbf{k}}-\phi _{b})$. Also, the
permittivity elements are not spatially dependent. In the wave equation, the
non-zero solution of $\mathbf{E}$ exists only if $\left\vert k_{0}^{2}%
\mathbf{\bar{\varepsilon}}_{r}-k^{2}\mathbf{\bar{I}}+\mathbf{kk}\right\vert
=0$. The determinant is a general relation for dispersion equation of the
bulk modes propagating in a gyrotropic medium in any arbitrary direction. By
plugging $\mathbf{\bar{\varepsilon}}_{r}$ tensor and $\mathbf{k}$ into the
dispersion relation, we obtain two solutions for $k_{z}$ as%
\begin{equation}
k_{zj}=\sqrt{-q_{\varphi }^{2}+\frac{1}{2\varepsilon _{t}}\left[ -\kappa \pm 
\sqrt{\kappa ^{2}-4\varepsilon _{t}\nu }\right] }
\end{equation}%
for $j\in \left\{ 1,2\right\} $ where%
\begin{eqnarray}
\kappa  &=&q_{\rho }^{2}(\varepsilon _{t}+\varepsilon _{a})+k_{0}^{2}\left(
\varepsilon _{g}^{2}-\varepsilon _{t}(\varepsilon _{t}+\varepsilon
_{a}\right)  \\
\nu  &=&\varepsilon _{a}\left( q_{\rho }^{2}-k_{0}^{2}\varepsilon
_{t}\right) ^{2}-\varepsilon _{a}\varepsilon _{g}^{2}k_{0}^{4}.
\end{eqnarray}%
Hence, the field in the gyrotropic region ($z<0$) can be written as a
superposition of two waves with the wave vectors $\mathbf{k}_{j}=\mathbf{k}%
_{tj}+q_{\rho }\mathbf{\hat{\rho}}$ where $\mathbf{k}_{tj}=$ $q_{\varphi }\mathbf{\hat{%
\phi}}+k_{zj}\mathbf{\hat{z}}$. The electric field vector in
the selected coordinate is written as $\mathbf{E}_{j}=\mathbf{E}_{0j}e^{i%
\mathbf{k}_{j}\cdot \mathbf{r}}=\left[ \alpha _{1}\mathbf{\hat{k}}%
_{tj}+\alpha _{2}\mathbf{\hat{\rho}}+\alpha _{3}\left( \mathbf{\hat{k}}%
_{tj}\times \mathbf{\hat{\rho}}\right) \right] e^{i\mathbf{k}_{j}\cdot 
\mathbf{r}}$. By plugging $\mathbf{E,}$ $\mathbf{k}$ and $\mathbf{\bar{%
\varepsilon}}_{r}$ into the wave equation, the unknown coefficients $\alpha
_{i}$ are obtained. Then we have 
\begin{equation}
\mathbf{E}_{0}\sim \mathbf{k}_{tj}+q_{\rho }\theta _{j}\mathbf{\hat{\rho}}%
+\Delta _{j}\left( \mathbf{k}_{tj}\times \mathbf{\hat{\rho}}\right) 
\end{equation}%
where%
\begin{equation}
\Delta _{j}\equiv \frac{i\varepsilon _{\text{g}}k_{0}^{2}}{\varepsilon _{%
\text{t}}k_{0}^{2}-k_{j}^{2}}\text{, \ \ \ }\theta _{j}\equiv \frac{%
-k_{tj}^{2}}{\varepsilon _{\text{a}}k_{0}^{2}-k_{tj}^{2}}
\end{equation}%
and the magnetic field is $\mathbf{H}=\left( \mathbf{k}\times \mathbf{E}%
\right) /\omega \mu _{0}$. In the isotropic region, the field can be
expanded as $\mathbf{E}=\left[ B_{1}\left( \mathbf{k}_{d}\times \mathbf{\hat{%
z}}\right) +B_{2}\mathbf{k}_{d}\times \left( \mathbf{k}_{d}\times \mathbf{%
\hat{z}}\right) \right] e^{i\mathbf{k}_{d}\cdot \mathbf{r}}$ where $\mathbf{k%
}_{d}=q_{\rho }\mathbf{\hat{\rho}}+q_{\varphi }\mathbf{\hat{\phi}}+%
k_{zd}\mathbf{\hat{z}}$ and $k_{d}=k_{0}^{2}\varepsilon _{r}$, taking into consideration the
equality of the tangential momentum at the interface. Finally, we decompose
the field vectors in both regions to their components in a regular polar
coordinate system. Let $\mathbf{\bar{Y}}_{g\text{ }}$ and $\mathbf{\bar{Y}}%
_{0}$ be the admittance tensors in the gyrotropic and the isotropic regions,
respectively. The tangential electric and magnetic field components are
related as%
\begin{equation}
\binom{-\eta _{0}H_{\varphi }}{\eta _{0}H_{\rho }}_{^{(2)}}=\mathbf{\bar{Y}}%
_{g\text{ }}\cdot \binom{E_{\rho }}{E_{\varphi }}_{^{(2)}}=\mathbf{\bar{Y}}%
_{g\text{ }}\cdot \mathbf{E}_{||}^{(2)}
\end{equation}%
and%
\begin{equation}
\binom{-\eta _{0}H_{\varphi }}{\eta _{0}H_{\rho }}_{^{(1)}}=\pm \mathbf{\bar{%
Y}}_{0\text{ }}\cdot \binom{E_{\rho }}{E_{\varphi }}_{^{(1)}}=\pm \mathbf{%
\bar{Y}}_{0\text{ }}\cdot \mathbf{E}_{||}^{(1)}
\end{equation}%
with the $\pm $ sign indicates upward and downward propagating waves
respectively and%
\begin{eqnarray}
\mathbf{\bar{Y}}_{0\text{ }} &=&\frac{1}{ik_{0}\gamma _{zd}}\left( 
\begin{array}{cc}
-\gamma _{zd}^{2}+q_{\rho }^{2} & q_{\rho }q_{\varphi } \\ 
q_{\rho }q_{\varphi } & -\gamma _{zd}^{2}+q_{\varphi }^{2}%
\end{array}%
\right)  \\
\mathbf{\bar{Y}}_{g\text{ }} &=&\frac{-1}{k_{0}\digamma }\left[ 
\begin{array}{cc}
\Lambda _{11} & \Lambda _{12} \\ 
\Lambda _{21} & \Lambda _{22}%
\end{array}%
\right] 
\end{eqnarray}%
have units of admittance where%
\begin{eqnarray}
\digamma  &=&q_{\rho }\theta _{1}(q_{\varphi }-i\Delta _{2}\gamma
_{z2})-q_{\rho }\theta _{2}(q_{\varphi }-i\Delta _{1}\gamma _{z1}) \\
\Lambda _{11} &=&q_{\rho }\Phi _{1}(q_{\varphi }-i\Delta _{2}\gamma
_{z2})-q_{\rho }\Phi _{2}(q_{\varphi }-i\Delta _{1}\gamma _{z1}) \\
\Lambda _{12} &=&-\theta _{2}\Phi _{1}q_{\rho }^{2}+\theta _{1}\Phi
_{2}q_{\rho }^{2} \\
\Lambda _{21} &=&\Delta _{1}k_{t1}^{2}(q_{\varphi }-i\Delta _{2}\gamma
_{z2})-\Delta _{2}k_{t2}^{2}(q_{\varphi }-i\Delta _{1}\gamma _{z1}) \\
\Lambda _{22} &=&-\theta _{2}\Delta _{1}k_{t1}^{2}q_{\rho }+\theta
_{1}\Delta _{2}k_{t2}^{2}q_{\rho }
\end{eqnarray}%
with $\Phi _{j}=\Delta _{j}q_{\varphi }\mathbf{-}i(\theta _{j}-1)\gamma _{zj}
$, $\gamma _{zd}=$ $-ik_{zd}$, and $\gamma _{zj}=ik_{zj}$. In the isotropic
region, $\mathbf{E}_{||}^{(1)}=\mathbf{E}_{||}^{p}+\mathbf{E}_{||}^{s}$ and $%
\mathbf{E}_{||}^{s}=\mathbf{\bar{R}(q)\cdot E}_{||}^{p}.$ By imposing the
continuity of tangential fields at $z=0$, the reflection tensor is $\mathbf{%
\bar{R}}=(\mathbf{\bar{Y}}_{0\text{ }}+\mathbf{\bar{Y}}_{g\text{ }%
})^{-1}\cdot (\mathbf{\bar{Y}}_{0}-\mathbf{\bar{Y}}_{g})$.

\section*{Acknowledgement}

Funding for this research was provided by the National Science Foundation
under grant number EFMA-1741673.

\section*{Data Availability}

The datasets generated during and/or analyzed during the current study are
available from the corresponding author on reasonable request.

\bibliographystyle{IEEEtran}
\bibliography{refrences.bib}

\end{document}